\documentstyle[prl,aps]{revtex}
\begin{document}
\draft

\twocolumn[\hsize\textwidth\columnwidth\hsize\csname@twocolumnfalse\endcsname

\title{Dynamics of $d$-wave Vortices: Angle-Dependent  Nonlinear Hall 
Effect}
\author{J. J. Vicente Alvarez, D. Dom\'{\i}nguez, and C. A. Balseiro}
\address{Centro At\'{o}mico Bariloche, 8400 S. C. de Bariloche,
Rio Negro, Argentina}
\date{\today}
\maketitle
\begin{abstract}
We study the dynamics of vortices in $d$-wave superconductors using a 
phenomenological time-dependent Ginzburg-Landau equation with mixing
of $s$ and $d$-wave components.
We present numerical simulations under an external driving 
current ${\bf J}$ oriented with an angle $\varphi$ with respect to the
$b$ crystal axis, calculating the vortex motion and  induced electric fields 
for $\kappa=\infty$. We find an intrinsic
Hall effect for $\varphi\neq 0$
which depends as $\sim\sin(4\varphi)$, and increases 
non-linearly with ${\bf J}$.

\end{abstract}


]                

\narrowtext

The discovery of high-$T_c$ superconductivity has renewed the interest
in the vortex state of type II superconductors \cite{reviews}. 
In conventional $s$-wave superconductors, the physics of the vortex
state was built upon two basic results: 
(i) vortices form a triangular lattice structure in equilibrium
as shown by Abrikosov \cite{abrikosov}, and 
(ii) when an external current is applied there is dissipation due 
to vortex motion \cite{BSNV}. 
There is now a growing experimental \cite{expd1,expd2} and 
theoretical \cite{theord} evidence
that the symmetry of the superconducting gap is most likely
to be d$_{x^2-y^2}$ wave. Therefore, it is now 
an issue of central interest how both the static and {\it dynamic}  
macroscopic properties of vortices are affected by the 
$d$-wave superconductivity.

The structure of a single vortex in a $d$-wave superconductor
was recently studied by  Volovik \cite{volovik}, Soininen {\it et al.}
\cite{soininen} and  Schopol and Maki \cite{maki}. It was found
in \cite{soininen}  that a $s$-wave component is
nucleated near the vortex core with opposite winding of the phase
relative to the $d$-wave component, and a four-lobe shape of the
amplitude. These results lead to formulate a phenomenological
Ginzburg-Landau (GL) theory with an order 
parameter with two components 
corresponding to the $s$-wave pairing
and the $d$-wave pairing respectively\cite{soininen,ren,berlinsky,vincent}. 
The main ingredient of this theory is a mixed gradient coupling
of the $s$ and $d$ components. This GL theory
showed that far away from the core the $s$ wave winding number
changes from $-1$ to $+3$ resulting in four additional nodes 
in the $s$-wave order parameter symmetrically placed along the
$\pm a$, $\pm b$ symmetry axes \cite{ren,berlinsky}. Using this approach,
Berlinsky {\it et al.} \cite{berlinsky} 
showed that, due to the fourfold anisotropy of individual
vortices,  the equilibrium lattice is  non-triangular at large fields, 
as seen in YBa$_2$Cu$3$O$_{7-\delta}$ \cite{expd2}.

The question we address in this Letter is how the four-lobe
structure of $d$-wave vortices affects the transport properties
of high-$T_c$ superconductors. We 
consider the {\em dynamics} of $d$-wave vortices 
when an external current ${\bf J}$ is applied.
The current is oriented with an angle $\varphi$ with respect to the
$b$ crystal axis, and we have calculated the induced electric field 
${\bf E}$ due to vortex motion \cite{MM}. 
This is represented  schematically in 
Fig.1(a), as well as  a possible experimental setup for the
measurement [Fig.1(b)].  

A vortex in motion is subject to Lorentz, viscous and hydrodynamic or Magnus  
forces which result in components both parallel and normal 
to its instantaneous velocity ${\bf v}_L$. This leads to transport
phenomena such as the Hall effect in the mixed state, 
which is still under active discussion 
\cite{BSNV,tdgl,dorsey,KK,AvO,harris}.  
One can distinguish three levels of treatment of this problem:
(i) {\it a macroscopic approach},
like the Bardeen-Stephen model \cite{BSNV};  
(ii) {\it an intermediate approach},
which considers  a time-dependent Ginzburg-Landau (TDGL)
equation; and 
(iii) {\it a microscopic approach} which deals with
the Bogoliubov-de Gennes equation (BdG) \cite{KK,AvO}. 
Up to now, all these approaches have been applied only to 
$s$-wave vortices. Here, we will study the dynamics of $d$-wave
vortices with a time-dependent version of the GL equation studied in 
\cite{soininen,ren,berlinsky}.

The GL free energy for a $d_{x^2-y^2}$ superconductor studied in
\cite{soininen,ren,berlinsky} has
an order parameter with two components $d({\bf r})$ and  $s({\bf r})$
and is given by: 
\begin{eqnarray}
f &=&\alpha_d|d|^2+\alpha _s|s|^2+\beta _d|d|^4 
+\beta _{sd}(s^{*2}d^2+s^2d^{*2}) \label{f} \\
&&+\gamma _d|{\bf\Pi}d|^2+\gamma _s|{\bf\Pi}s|^2 +  \nonumber \\
&&+\gamma_{v}[(\Pi _bs)^{*}(\Pi _bd)-(\Pi _as)^{*}(\Pi _ad)+c.c.]+h^2/8\pi
\nonumber
\end{eqnarray}
where ${\bf\Pi}=(-i{\bf\nabla}-e^{*}{\bf A}/\hbar c)$ is defined
in the $ab$ plane, ${\bf A}$ is the vector potential and 
${\bf h}={\bf\nabla }\times {\bf A}$. 
The critical temperature for $d$-wave superconductivity $T_c^d$ is assumed to
be larger than the one corresponding to $s$-wave pairing, 
{\it i.e.} $\alpha _d<0$ and $\alpha_d<\alpha _s$, and 
all the other constants are taken to be positive. 
In \cite{ren} it was obtained the relation: 
$\gamma _d=\gamma_{v}=\gamma_s/2$.
For generality, we consider these constants to be independent 
with the constrain $\gamma _d\gamma _s>\gamma_{v}^2$.

We assume that  the TDGL equations of motion for 
the order parameters
$d$ and $s$ are simply given as \cite{comment1} 
\begin{equation}
\begin{array}{ll}
\hbar \left[ \partial _t +i\frac{e^{*}}\hbar \phi \right] d=
-\Gamma \frac{\partial f}{\partial d^{*}}, 
& \hbar \left[ \partial _t +i\frac{e^{*}}\hbar \phi \right] s=
-\Gamma \frac{\partial f}{\partial s^{*}},
\end{array}
\label{dtdgl}
\end{equation}
with $\Gamma$ the relaxation parameter and $\phi$ the scalar potential. 
We consider strongly type-II materials in the
$\kappa =\infty$ limit. Therefore, the magnetic
field is uniform and time-independent,
and we consider the case ${\bf B}\parallel{\bf c}$.
The system of equations is closed after considering  current
conservation: ${\bf \nabla}\cdot{\bf J}=0$, with
the total current density ${\bf J}={\bf J}_s+{\bf J}_n$.
The supercurrent ${\bf J}_s$ is given by 
\begin{eqnarray}
{\bf J}_s &=&\frac{2e}{\hbar}
\gamma _d\left[ d^{*}\left( {\bf\Pi }d\right)\right] + \frac{2e}{\hbar}
\gamma _s\left[ s^{*}\left( {\bf\Pi }s\right)\right] \label{GLe} \\
&&+\frac{2e}{\hbar}
\gamma_{v}\left\{ \left[ s^{*}\left( \Pi _bd\right) +\left( \Pi
_bs\right) ^{*}d\right] {\bf\hat{b}}\right. \nonumber \\
&&\left.-\left[ s^{*}\left( \Pi
_ad\right) +\left( \Pi _as\right) ^{*}d\right] {\bf\hat{a}}\right\}+c.c. 
\nonumber
\end{eqnarray}
while the normal current is ${\bf J}_n=\sigma_n.{\bf E}$,
with $\sigma_n$ an effective normal-state conductivity. 

These TDGL equations are  solved numerically using a 
finite-difference algorithm \cite{met} in a square grid along the
$x,y$ axis with discretization $\Delta x=\Delta y = 0.2 \xi$ and
system size $25.6\xi\times25.6\xi$, with
$\xi=\sqrt{\gamma_d/|\alpha_d|}$. 
We take periodic boundary conditions for the physical
quantities ${\bf h},{\bf E},{\bf J}_s,|d|$ and $|s|$. 
One convenient gauge choice is a time independent ${\bf A}=n_vB_0y
{\bf\hat{x}}$, where $n_v$ is the number of vortices in the sample, 
$B_0=\Phi_0/S$ and $\Phi _0$ is the flux quantum, 
and ${\bf E}=-{\bf\nabla }\phi $. 
The scalar potential $\phi$  can be decomposed as $\phi 
({\bf r})=\phi _0({\bf r})-
{\bf E}_0(t)\cdot{\bf r}$ with $\phi _0({\bf r})$ being periodic 
and satisfying 
$\nabla ^2\phi _0=\sigma_n^{-1}{\bf\nabla}\cdot{\bf J}_s$ and 
${\bf E}_0(t)=\sigma_n^{-1}\left[{\bf J}_{ext}-1/S\int d^2r\;
{\bf J}_s(t)\right]$, since for an applied external current, 
from global current conservation, we have 
${\bf J}_{ext}=1/S\int d^2r\;{\bf J}$. 
To ensure the periodicity
of ${\bf J}_s$ the phases of $d$ and $s$ must obey the boundary conditions: 
$\theta ({\bf r}+L_x{\bf\hat{x}})=\theta ({\bf r})+\Theta_x$ and 
$\theta ({\bf r}+L_y{\bf\hat{y}})=\theta ({\bf r})+gx+\Theta _y$ 
where $g=n_vB_0L_y$, $\partial _t\Theta _x=L_xE_{ox}(t)$ and  
$\partial _t\Theta _y=L_yE_{oy}(t)$.
The external current is taken to be along the discretization $y$-axis,
and the angle $\varphi$ between $\hat{a}$ and $\hat{x}$ [see Fig.1(a)]
is varied by ``rotating'' the GL equations. (We obtain
similar results by fixing the numerical grid to $(x,y)=(a,b)$ and
rotating ${\bf J}_{ext}$, but the effects of the discretization are 
stronger in this case). The equation are integrated with a time
step $\Delta t = 0.001\hbar/|\alpha_d|\Gamma$ averaging
over $5000$ steps after a transient of $10^4$ steps.

A vortex moving with a velocity ${\bf v}_L$ induces an average electric field
$\langle{\bf E}\rangle=-{\bf v}_L\times\langle{\bf B}\rangle$
\cite{tdgl}. Since from current conservation we have that
${\bf\nabla}\cdot{\bf E}=-\sigma_n^{-1}{\bf\nabla}
\cdot{\bf J}_s=4\pi\rho$, the exchange of supercurrents into normal currents
occurring in the vortex core results in a dipolar charge distribution 
with momentum ${\bf p}\perp{\bf v}_L$, 
such that $\langle{\bf E}\rangle\propto{\bf p}$. 
Under an external current ${\bf J}_{ext}$
there is a Hall effect when the current is not parallel to 
$\langle{\bf E}\rangle$, defining a Hall angle $\tan\theta_H=
\langle E_\perp\rangle/\langle E_\parallel \rangle$. 
The TDGL equation has been used in the past to model vortex dynamics
in  conventional superconductors \cite{tdgl,dorsey}. In this case,
it has been shown by Dorsey \cite{dorsey} that a Hall effect can be
induced from two mechanisms: (i) by a complex relaxation parameter
$\Gamma=\Gamma_1+i\Gamma_2$ with $\Gamma_2\not=0$ (modeling a
hydrodynamic contribution to the Hall effect), and  (ii) by
a normal state off-diagonal conductivity $\sigma_{n,xy}$ (modeling
a quasiparticle core contribution). We consider here
the TDGL equations (\ref{dtdgl}) with $\Gamma_2=0$ and
$\sigma_{n,xy}=0$, to look for additional contributions to
the Hall effect coming from the lack of rotational symmetry in 
$d$-wave superconductivity.

Let us first consider the motion of a single $d$-wave vortex under a 
current  $J_{ext}=0.6J_c$ with
$J_c=\frac{2e|\alpha_d|\gamma_d}{\hbar3\sqrt{3}\beta_d\xi_d}$ 
(the critical current of the pure $d$-wave solution),
and parameter values $\alpha_s/|\alpha_d|=1.25,
\gamma_v/\gamma_d=1, \gamma_s/\gamma_d=2, \beta_{sd}/\beta_d=0.5, 
\Gamma=1, \sigma_n=(4e^2/\hbar)(\gamma_d/2\beta_d\Gamma)$, 
and for different values of  $\varphi$, see Figure 2.
We find that when ${\bf J}_{ext}$ is in  one of the directions of
maximum symmetry ($\varphi=0^{o}, 45^{o}, 90^{o}$) the vortex moves 
with ${\bf v}_L\perp{\bf J}_{ext}$, and therefore $\theta_H=0$. 
This is shown in Fig.2(a-c) for $\varphi=0$.
We see in Fig.2(a)  that the amplitude $|d|$ has a square-like
shape aligned with the $ab$ axes as seen in the static case
\cite{berlinsky}. We also see that  $|d|$ is depressed in the
forward direction of motion where the vortex currents add to
${\bf J}_{ext}$. The corresponding $s$-wave amplitude $|s|$ shows the
four-lobe structure corresponding to the four satellite vortices 
\cite{berlinsky}, but with a strong deformation due to the vortex motion. 
Also the relative position of the satellite vortices is displaced 
backwards with respect to the static vortex. 
Furthermore, there is a finite $|s|$ induced at
long distances because even in the absence of vortices a
current ${\bf J}_{ext}$ induces a $s$-wave component
\cite{ruso}.The motion induced charge 
$\rho({\bf r})=\frac{-1}{4\pi\sigma_n}{\bf\nabla}\cdot{\bf J}_s$ is 
shown in Fig.2(c). It corresponds to a dipole 
${\bf p}\propto{\bf E}_{av}$ in the direction of ${\bf J}_{ext}$, 
and therefore $\theta_H=0$. 
The $d$-wave state gives only
a small additional quadrupolar contribution to ${\bf E}$ in this case.
On the contrary, when the current ${\bf J}_{ext}$ is not in a
direction of symmetry there is always a finite Hall effect, which is maximum
for $\varphi=22.5^{o}$, the case shown in Fig.2(d-f). In Fig.2(d) and (e) we
see that the amplitudes $|d|$  and $|s|$ show now a motion-induced
deformation that is determined both by the direction of ${\bf
J}_{ext}$ and by the orientation of the $ab$ axes (i.e. amplitude
depression in the direction perpendicular to ${\bf J}_{ext}$ and
four-lobe  structure in the $ab$ orientation). This
results in vortex motion with a velocity ${\bf v}_L$ in a direction
that is not perpendicular to  ${\bf J}_{ext}$. 
In Fig.2(f), we see from the plot of $\rho({\bf r})$ 
that the induced dipole ${\bf p}$ is not collinear with  ${\bf J}_{ext}$. 
Therefore, there is  a transverse component in 
$\langle{\bf E}\rangle$ giving $\theta_H\not=0$.
 
In general, we find that both components of the electric field, 
longitudinal and transversal to the current,  oscillate with
$\varphi$.  We find that this results in a Hall angle  with
a dependence of $\tan\theta_H \propto \sin(4\varphi)$.
In Fig.~3(a) we show the dependence $\tan\theta_H$ 
with the angle $\varphi$ of the current for
$\gamma_v/\gamma_d=1, \gamma_s/\gamma_d=2, \alpha_s/|\alpha_d|=5, 
\beta_{sd}/\beta_d=0.5$ and $J_{ext}\equiv J=0.6J_c$. 
In conventional superconductors, $\tan\theta_H$ should be current
independent for small $J/J_c$. In Fig.~3(b) we show the current
dependence of $\tan\theta_H$ for $\varphi=22.5^o$. We
see that the Hall angle  depends with the external current as  
$\tan\theta_H \propto (J/J_c)^2$. 
This means that the Hall effect vanishes in linear
response theory when $\Gamma_2=0, \sigma_{n,xy}=0$, as
shown by Dorsey \cite{dorsey}. In a rotationally symmetric system,
it should also vanish for all orders of $J/J_c$. 
Here, we find that in the $d$-wave case there is a strong contribution
to the  Hall effect in the non-linear regime.
We have also added an imaginary component to the dissipation parameter, 
$\Gamma_2\not=0$. We have chosen $\Gamma_2$ such
to give a finite $\tan\theta_H$ for $\varphi=0$ similar to
the typical experimental values \cite{harris}.
We see in the insets of Figures 3(a) and (b) 
that the $d$-wave induced Hall effect is additive to the  Hall angle
induced by $\Gamma_2$. 
We have also studied  the motion of a vortex lattice for different
values of the field $B\xi^2/\Phi_0=0.007,0.021,0.063$, 
and we find $\tan\theta_H$ is nearly field independent \cite{largo}. 

A simple understanding of this angle-dependent
non-linear Hall effect can be obtained using the generalized
London theory of Affleck {\it et al} \cite{affleck}. The supercurrent
${\bf J}_s$ can be written in terms of the superfluid velocity ${\bf
v}_s \equiv {\bf v} = \nabla\theta_d -(2e/hc){\bf A}$ as:
\begin{eqnarray}
{\bf J}_s&=&\frac{2e}{\hbar} g^2\gamma_d\left\{{\bf v} -2\epsilon\xi_d^2
\left[(\hat{\bf b}v_b -\hat{\bf a}v_a)(v_b^2-v_a^2)\right.\right.\nonumber \\
&&\left.\left.-(\hat{\bf b}\partial_b -\hat{\bf a}\partial_a)
(\partial_bv_b-\partial_av_a)\right]\right\},
\label{london}
\end{eqnarray}
with $g^2=|\alpha_d|/2\beta_d$ and the small parameter 
$\epsilon=(|\alpha_d|/\alpha_s)(\gamma_v/\gamma_d)^2$. The main
result here is that the ${\bf v}_s$ is not collinear ${\bf J}_s$, 
except when oriented along the directions of  symmetry of
the $ab$ crystal. In a purely dissipative dynamics
the balance of forces acting on a vortex line is
given as $\eta {\bf v}_L = 2en_s \Phi_0{\bf v}_s\times\hat{\bf
n}$, with $\eta$ a dissipation parameter (determined
by $\sigma_n$ and $\Gamma_1$ \cite{tdgl,dorsey})
and $\hat{\bf n}$ the direction of ${\bf B}$. Here 
${\bf v}_s$  is the supervelocity far away from the vortex core induced by 
the driving current ${\bf J}_{ext}\equiv {\bf J}$. From (\ref{london})
we obtain that the external force acting on the vortex line 
is, at first order in $\epsilon$,
\begin{equation}
{\bf f}_L \approx \frac{\Phi_0}{c}{\bf J}\times\hat{\bf n} + 
 \frac{\Phi_0}{c}\frac{\hbar}{2e}
\frac{2\epsilon\xi^2}{g^2\gamma_d}(J_b^2-J_a^2)
(\hat{\bf b}J_b - \hat{\bf a}J_a)\times\hat{\bf n}.
\end{equation} 
Therefore, the resulting ``Lorentz force'' is not perpendicular to ${\bf J}$.
Since the motion induced electric field is 
${\bf E}={\bf v}_L\times{\bf B}$, we obtain a Hall angle given as
\begin{equation}
\tan\theta_H = A \epsilon \left(\frac{J}{J_c}\right)^2 \sin4\varphi,
\label{halle}
\end{equation}
with $A$ a constant  of the order of unity. Even when this 
analysis is very crude, we have obtained the
same result by a generalization of the Bardeen-Stephen model 
within a perturbative treatment of the London
theory of Affleck {\it et al.} in the small parameter $\epsilon$ \cite{largo}.
We have verified  the Eq.(\ref{halle}) by fitting our numerical results
of Fig.~3. In Fig.3(a) we have fitted the $\sin4\varphi$ dependence
and in Fig.3(b) we have fitted the $(J/J_c)^2$ dependence for low
currents. In Fig.3(c) we show the dependence with
$\epsilon=(\alpha_d/\alpha_s)(\gamma_v/\gamma_d)^2$ for different
values of $\alpha_d/\alpha_s$ and $\gamma_v/\gamma_d$ taken
independently. We see that for small $\epsilon$ this is an 
adequate parameter combination and that the Hall effect is linear in 
$\epsilon$. From our numerical fits we obtain $A\approx0.28$.
Therefore, a measurement of the angle and current dependence of 
$\tan\theta_H$ will give a direct estimate of the parameters
of the GL model of Eq.(\ref{GLe}).

In conclusion, we have shown that in a $d$-wave superconductor,
the four-lobe structure of vortices leads to a  non-linear angle
dependent Hall effect. 
The effect can be explained from the 
fact that the supercurrent and the supervelocity are not
collinear as shown in the generalized London model of \cite{affleck}.
Therefore, this result does not depend on the  detail of the quasiparticle
physics in the vortex core, which is not fully described in a TDGL
treatment. Instead, this effect is clearly a consequence of the breaking of
rotational  invariance in $d$-wave superconductivity.
Close to $T_c$, in mean field,
the depairing current vanishes as $J_c \sim
t^{3/2}$, with $t=|T_c-T|/T_c$, and $\epsilon \sim t$,
then we have that $\tan\theta_H \sim t^{-2}$ at constant $J$. Therefore,
the effect should be more noticeable close to $T_c$, where the
nonlinearity in currents is  stronger. An experimental setup 
as  shown schematically in the Fig.1(b)
would allow a measurement of this angle-dependent Hall effect in an
untwinned crystal. As we have found,
one will have $\theta_H=\theta_H^0+\theta_H^d(\varphi)$.
Typical experiments in YBa$_2$Cu$_3$O$_{6+y}$ \cite{harris}, 
which are at $\varphi=0$, give  
$\tan\theta_H^0\approx 0.05$ close to $T_c$.
For reasonable values of the parameters of the GL theory \cite{ren}
we have $\epsilon\approx 0.2$, which for $J/J_c=0.5$ and
$\varphi=22.5^o$ Eq.~(\ref{halle}) gives $\tan\theta_H^d\approx
0.02$,  therefore a dependence with $\varphi$ could be observable.  
In principle,
the anisotropy between the $a$ and $b$ axis should also be considered,
but this can lead only to dependences $\sim \sin2\varphi$, at most.
Therefore, we predict that a measurement of a Hall effect with
an angular dependence $\sim \sin4\phi$ would be a direct
evidence of the $d$-wave character of the High-$T_c$ superconductivity
in the transport properties of vortices.

We acknowledge CONICET for financial support 
and we also acknowledge discussions with 
L. N. Bulaevskii, F. de la Cruz, E. Osquiguil and
G. Buscaglia.

\begin{figure}
\caption{Schematic view of (a) the vortex with the crystalline axis
and the frame of reference (see text) and (b) the experimental setup
proposed to measure the angle dependent Hall effect.}
\end{figure}

\begin{figure}
\caption{
Contour plot of the amplitudes of d-wave ((a) and (d)) and s-wave
((b) and (e)) components of the order parameter of a moving
vortex for a current $J_{ext}=0.6J_c$. 
The panels (c) (f) show the contour plot of the induced
electric charge. The parameters are $\frac 1 2 \gamma_{s} =\gamma_{d} 
=\gamma_{v}$,$\alpha_{s}=1.25 |\alpha_{d}| $ and
$\beta_{d}=2\beta_{sd}$. In (a), (b) and (c) the angle between 
$J_{ext}$ and the crystal $b$ axis is $\varphi=0$  
 and in (d),(e) and (f) is $\varphi=22.5^o$.}
\end{figure}

\begin{figure}
\caption{Plot of the Hall angle $\tan\theta_H$ (a) as function of the angle 
of the current $\varphi$ for $J_{ext}=0.6J_c$, (b)
as a function of $(J_{ext}/J_c)^2$ for $\varphi=22.5^o$. The
parameters are $\frac 1 2 \gamma_s = \gamma_d = \gamma_{v}$,
$\alpha_{s}=5|\alpha_{d}|$ and $\beta_{d}=2\beta_{sd}$.
The continuous lines are the fittings with Eq.(6). In the insets the 
squares, circles and triangles correspond to $\Gamma_2/\Gamma_1
=0,-0.02, -0.3$. In (c) $\tan\theta_H$ as a function of
$\epsilon=(\gamma_{v}/\gamma_d)^2|\alpha_d|/ \alpha_s $ 
for $J_{ext}=0.6 J_c$ and $\varphi=22.5^o$.}
\end{figure}


\begin{references}
\bibitem{comment1} In general $\Gamma$ could be a tensor with components
$\Gamma_{ss}, \Gamma_{sd}, \Gamma_{dd}$. We consider here 21$\Gamma$ scalar for
simplicity.
\end{references}
\end{document}